# The USA is an indisputable world leader in basic medical and biotechnological research

Ricardo Brito[1]*, Alonso Rodríguez-Navarro[1,2]

[1] *Departamento de Estructura de la Materia, Física Térmica y Electrónica and GISC, Universidad Complutense de Madrid, Plaza de las Ciencias 3, 28040, Madrid, Spain*

[2] *Departamento de Biotecnología-Biología Vegetal, Universidad Politécnica de Madrid, Avenida Puerta de Hierro 2, 28040, Madrid, Spain*

\* Corresponding author: Ricardo Brito;  *e-mail address*: brito@ucm.es

A country's research success can be assessed from the power law function that links country and world rank numbers when publications are ordered by their number of citations; a similar function describes the distribution of country papers in world percentiles. These functions allow calculating the $e_p$ index and the probability of publishing highly cited papers, which measure the efficiency of the research system and the ability of achieving important discoveries or scientific breakthroughs, respectively. The aim of this paper was to use these metrics and other parameters derived from the percentile-based power law function to investigate research success in the USA, the EU, and other countries in hot medical, biochemical, and biotechnological topics. The results show that, in the investigated fields, the USA is scientifically ahead of all countries and that its research is likely to produce approximately 80% of the important global breakthroughs in the research topics investigated in this study. EU research has maintained a constant weak position with reference to USA research over the last 30 years.

# Introduction

Countries and institutions require reliable research assessment methods to determine the profitability of their research investments. In the absence of reliable research assessments, the actual economic and societal benefits of research and its contribution to the progress of knowledge cannot be judged. Research produces knowledge and, in most cases, this knowledge is used by other systems to produce economic and other societal benefits (Bornmann and Marx 2014); however, the production of knowledge depends on the *breakthrough potential* of the research system and is not proportional to the amount of R&D investment and capital as assumed in most econometric analyses (Rodríguez-Navarro and Brito 2018). The inaccuracy of this assumption explains that, although the economic benefits of research are generally widely accepted, "no simple model of the nature of the economic benefits from basic research is possible" (Salter and Martin 2001, p. 527). Consequently, in the absence of accurate assessments, the low benefits of research in lowly performing research systems could be assumed to occur because the benefits of research are intrinsically low. This conclusion could lead to the political response of decreasing research funding instead of improving research performance.

Although evaluating the performance of research systems in terms of knowledge production is highly required, many issues make it difficult. Within the scope of this study, the most important issue is the intangible nature of knowledge progress (Martin and Irvine 1983), which is the product that should be measured. Publications are tangible and counting them is widely used in research assessments as a substitute for measuring knowledge production. However, this substitution is formally incorrect as some publications are unnecessary and others report data that are interesting for researchers but irrelevant for society at large. In fact, the publications that report discoveries and breakthroughs are a small amount, but are supported by a large number of "normal publications," adopting a Kuhnian terminology (Kuhn 1970). In the end, the ratio between the number of scientific publications that report discoveries or breakthroughs and the total number of publications is very low (Rodríguez-Navarro 2012) and is expected to vary across countries. Therefore, assessing the share of breakthrough publications by countries and institutions is impossible through the counting of these publications; for example, if a research field with 100,000 annual publications produces only 20–30 real discoveries or breakthroughs, it will take most small countries or institutions several years to produce one of them, and the number of years will be highly variable even across countries with similar number of publications.

Therefore, counting publications, either the total number, which is irrelevant, or those reporting breakthroughs, which is impossible in most cases, cannot be general methods of research assessment. A likely solution to this conundrum would be to calculate the cumulative probability of producing breakthrough publications. Multiplying this probability by the total number of publications produces the cumulative frequency of breakthrough publications for a country or institution.

This evaluation procedure based on a mathematical probability is possible because citation distributions follow lognormal functions that allow calculating cumulative probabilities at any citation level (Rodríguez-Navarro and Brito 2019). However, a simpler method is by utilizing the percentile-based double rank approach, which allows calculating the $e_p$ index. Percentile frequencies obey power laws, and the $e_p$ index is a derivative of the exponent of these power laws (Brito and Rodríguez-Navarro 2018; Rodríguez-Navarro and Brito 2019). The $e_p$ index increases or decreases depending on whether the frequency of country publications among global publications increases or decreases when going up the citation scale; a uniform distribution of country publications implies an $e_p$ index value of 0.10. The probability that a random paper from a country reaches a top percentile $x$ is calculated by simply raising $e_p$ to a power, as explained below, which implies that the $e_p$ index is a measure of the *intrinsic efficiency* or *breakthrough potential* of the research system (Rodríguez-Navarro and Brito, 2018)

In a previous study, this approach was used to assess the performance of technological research in three world geographical areas: the USA, the EU, and the rest of the world (Rodríguez-Navarro and Brito 2018). The results showed that the USA and some non-USA, non-EU countries dominate at similar levels the global production of breakthrough papers in fast evolving research topics while the EU lags far behind. In low evolving technological topics research is dominated by non-USA, non-EU countries and the USA and the EU lag slightly behind. However, this study only considered physical- and chemical-based technologies. Although these technologies support a large proportion of economic growth, two biological research fields, namely medicine and biotechnology, have a large societal relevance. Therefore, the present study is focused on these important research fields and aims to find out how the progress at the forefront of knowledge in these areas is globally distributed.

This progress at the forefront of knowledge is related to radical innovations, but incremental innovations, which are not based on important scientific breakthroughs, also support technological progress (Dewar and Dutton 1986); in the case of medicine, research is even more diverse (Röhrig et al. 2009). Although the probabilities of achieving scientific breakthroughs are only related to radical changes, the $e_p$ index that we also study here has a broader meaning because it reveals the *intrinsic efficiency* of the research system (Rodríguez-Navarro and Brito 2018).

## Previous considerations and methods

**Research performance parameters**

As in previous studies, using the percentile-base double rank approach we calculated three evaluation parameters for each country or group of countries: the $e_p$ index, the probability that a country or institution's publication has to reach the top 0.01 citation percentile, and the cumulative frequency of publications in this percentile.

The calculation of these parameters was described previously (Brito and Rodríguez-Navarro 2018; Rodríguez-Navarro and Brito 2018, 2019). Briefly, global and country publications were ranked in parallel using the number of citations in decreasing order; the percentile limits in the global list (any percentile from 1% to 100%) were fixed in according to the rank numbers and turned into the number of citations of the last paper of the selected percentile. Then, the country number of publications in each selected percentile was equal to the number of papers with the same or higher number of citations as the last paper of the selected percentile in the global list (i.e., the country ranking number of the last paper). When in the global and country lists the percentile limits occurred in sets of publications with the same number of citations (Waltman and Schreiber 2013), the number of tied publications in the country set was fixed using a proportional method (i.e., proportional number of tied publications in global and country lists). This method is very accurate and can be used for those researches that cannot download hundreds of thousands of publications. Figure 1 shows the accuracy of this method in biological areas, which is similar to that found in technological areas (Figure 1 in Rodríguez-Navarro and Brito 2018).

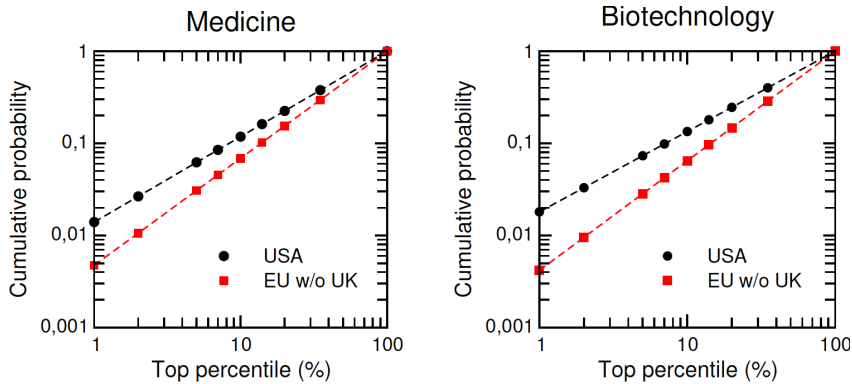

*Figure 1. Cumulative probability plots for a paper from the USA and the EU excluding the UK to reach a specific top percentile. Publications in 2014.*

After counting the number of publications in a series of percentiles, the percentile distribution of the number of publications (i.e., cumulative frequencies) was fitted to a power law as shown in Figure 1. The $e_p$ index is equal to 10 raised to minus the exponent of the fitted power law function (Rodríguez-Navarro and Brito 2019). The equations are:

$$e_p = 10^{-\alpha} \qquad (1)$$

where $\alpha$ is the exponent of the power law function. The cumulative probability was calculated from the function:

$$P(x) = e_p^{(2 - \lg x)} \qquad (2)$$

where $x$ is the selected percentile. In the present study, the top 0.01 percentile was selected for the reasons given elsewhere (Rodríguez-Navarro and Brito 2018). The expected frequency of papers in the top 0.01% of the most cited publications is equal to the cumulative probability, $P(x)$, multiplied by the number of publications. This cumulative frequency is named $P_{\text{top }0.01\%}$ using the Leiden Ranking notation (Bornmann et al. 2015; Waltman et al. 2012).

It is worth noting that the three parameters are highly related but different. The $e_p$ index is both percentile- and size-independent, and reveals the *intrinsic efficiency* of the research system (Rodríguez-Navarro and Brito 2018). The other two parameters depend on the selected percentile; the probability is size independent and the cumulative frequency is size dependent.

**Topics**

The purpose of this study was to investigate the country's research performances across the world in basic biology and its application to biotechnology and medicine. To perform it we used the Web of Science Core Collection Advanced Search feature; to select the terms of the search query we took into consideration that country's research performance varies enormously not only across research fields (e.g., physics, chemistry, or biology), which seems logical, but also depending on the research activity of the field, (i.e., hot or quiescent topics; Bonaccorsi 2007; Brito and Rodríguez-Navarro 2018; Rodriguez-Navarro and Narin 2018; Sachwald 2015). Therefore, we focused in research areas and topics that are currently highly investigated.

For basic biology and biotechnology, the selection of the query terms was simple and could be restricted to two Research Areas of the Web of Science Core Collection available for the Advanced Search (SU=): Biochemistry & Molecular Biology and Biotechnology & Applied Microbiology. The adequacy of these research areas to our purpose was clear for the scientific coincidence and because they include very highly cited research topics. In fact, among the 22 Research Fields of the Essential Science Indicators, the field Molecular Biology and Genetics, which is in the basis of the selected Research Areas, had the highest number of citations per paper of the 22 Research Fields.

The construction of the search query for basic medical research required more consideration because none of the Research Areas that can be included in the Advanced Search query fulfilled our requierements. Therefore, we constructed the query with hot topics (TS=) in basic medical research; to select these topics we attended to two obviuos conditions: they had to be basic and highly cited topics. To find these topics, firstly we retrieved the papers published in *Nature*, *Science*, and the *Proceedings of the National Academy of Sciences USA* and order them by their number of citations. Then we selected the most frequent biological topics that fulfilled our conditions and checked, one by one, that they were highly cited among global publications including all journals. From the results we selected: cancer, microbiota, stem cell*, immunity, and inflamma* (the asteric denotes truncation). Although these topics do not include all possible highly cited topics in basic medical research they are the most important and a representative sample of this type of research. In fact, their number of publications was high, 179,951 in 2018, almost the same number than in the whole research area of Chemistry, 190,290. The annual number of publications on individual topics was different; cancer and inflamma* represente almost 60% and 30%, respectively, of the total number of publications. The topic CRISPR was identified only in recent years; highly cited papers were published starting in 2007 and they were frequent since 2013. Although the effect of this topic was restricted to the last year of our study, we included it for its current high biological importance and certain increas of significance in future studies.

**Countries**

We divided the world into three geographical research areas: the ERA (European Research Area), the USA, and Others (i.e., all countries excluding the ERA countries and the USA; all countries means the 50 most productive countries in 2014). These areas were analyzed independently, omitting collaborative publications between them. The eight largest and scientifically most active countries in the ERA: Germany, France, UK, Italy, Spain, Sweden, The Netherlands, and Switzerland were also studied. In Others, we selected the six most active countries in the selected research topics and areas: Australia, Canada, China, Japan, South Korea, and Taiwan. In some cases instead of the ERA countries we record results for the EU excluding the UK in order to cancel out the dominant role that Switzerland and the UK play in ERA research (Rodríguez-Navarro and Brito 2018).

In the 15 selected countries we retrieved only domestic papers (all authors in the same country or set of countries; see discussion below). Some collaborative publications between two countries were also studied to complement the study of domestic papers.

**Bibliometric searches**

Bibliometric searches were performed in the Science Citation Index Expanded of the Web of Science Core Collection (WoS), using the "Advanced Search" feature. For highly cited medical topics we used: TS=((cancer OR crispr* OR microbiota OR stem cell* OR immunity OR inflamma*) NOT (statistics OR trial OR survey)), and for biochemistry and biotechnology we used: SU=((biochemistry & molecular biology OR biotechnology & applied biotechnology OR cell biology OR microbiology) NOT (computer science OR mathematical & computational biology)) NOT TS=(cancer OR crispr* OR microbiota OR stem cell* OR immunity OR inflamma* OR statistics OR trial OR survey). Although in our analysis very highly cited papers are not used for fitting the power law, clinical-trials and statistical papers, which are not breakthroughs but are normally highly cited, were specifically excluded because they might have influence at all citation levels. We retrieved only "Articles," which excludes review papers, because in many cases review papers receive more citations than the original articles in which they are based. Searches were performed between February 23 and March 5, 2018. Some countries were analyzed in different days but each analysis in a different day was complete including world and country citation distributions.

## Results

**Basic medical research**

To obtain a first overview of world research in hot medical topics, we studied the evolution of the number of papers and the three aforementioned parameters—$e_p$ index, probability that a random papers reaches the to 0.01 percentile, and the $P_{top\,0.01\%}$—between 1984 and 2014 (Figure 2). The number of publications in the selected medical topics increased enormously over these 30 years, from 3,686 to 148,375 annual publications. Growth in the USA and the ERA was similar, while in Others it was much higher, especially over the last 10 years (2004–2014). Throughout this time, the research performance in the USA, as revealed by the $e_p$ index and the paper probability of reaching the top 0.01 percentile, was much higher than in the ERA and Others. The performance of the USA was already the best in 1984 and it increased over time. The $e_p$ index showed a clear tendency of the ERA catching up with the USA in the 1984–1994 period; however, this tendency decreased in the 1994–2004 period and disappeared in 2004–2014. Even in 1984–1994, the paper probability of reaching the top 0.01 percentile showed that the ERA was clearly lagging compared to the USA. This apparently contradictory trend between the $e_p$ index and paper probability of reaching the top 0.01 percentile was due to the original great difference between the USA and the ERA, and the mathematical relationship between both parameters, which is exponential. Because of the original large difference, the ERA should increase its $e_p$ index more than the USA to decrease the difference in paper probability of reaching the top 0.01 percentile; for example, if the USA ($e_p \approx 0.10$) increased the $e_p$ index by 10% the ERA ($e_p \approx 0.06$) should increase it by 25% to similarly increase the probability.

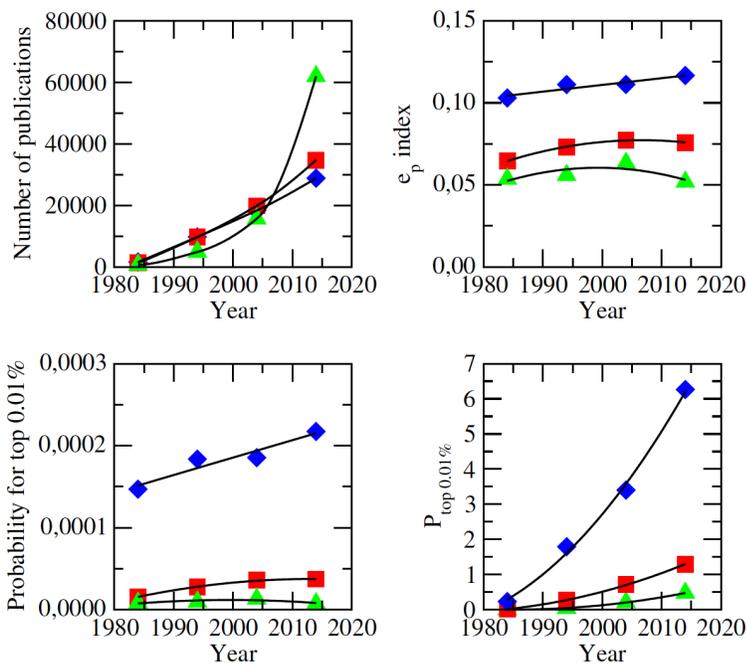

Figure 2. Evolution of research performance in the USA, the European Research Area (ERA), and other countries in top medical topics over 30 years. Curves are drawn to guide the eye. Symbols: squares, the USA; diamonds, the ERA; triangles, other countries.

Due to the large increase in the number of publications, the cumulative frequency of publications in the top 0.01 percentile ($P_{top\,0.01\%}$) increased in the three geographical areas studied here; however, because of the higher US research performance, the difference between the USA and both the ERA and Others increased permanently over the 30-year period.

Table 1. Research performance parameters in hot medical topics, in 15 selected cases. Publications in 2014; domestic counts.

| Countries | Number of publications | $e_p$ index | Paper probability for top 0.01% | $P_{(top\,0.01\%)}$ |
|---|---|---|---|---|
| Switzerland | 734 | 0.1479 | 0.0004782 | 0.3510 |
| USA | 28818 | 0.1180 | 0.0001936 | 5.5778 |
| UK | 3398 | 0.0822 | 0.0000456 | 0.1550 |
| Netherlands | 1721 | 0.0814 | 0.0000439 | 0.0755 |
| Germany | 4817 | 0.0806 | 0.0000423 | 0.2038 |
| Sweden | 1094 | 0.0791 | 0.0000392 | 0.0429 |
| France | 3175 | 0.0721 | 0.0000270 | 0.0856 |
| Canada | 2705 | 0.0670 | 0.0000202 | 0.0546 |
| EU w/o UK | 27241 | 0.0665 | 0.0000196 | 0.5331 |
| Australia | 2155 | 0.0644 | 0.0000172 | 0.0370 |
| Spain | 2419 | 0.0579 | 0.0000113 | 0.0272 |
| China | 23602 | 0.0566 | 0.0000103 | 0.2426 |
| Italy | 4231 | 0.0543 | 0.0000087 | 0.0369 |
| South Korea | 5751 | 0.0484 | 0.0000055 | 0.0315 |
| Japan | 7925 | 0.0470 | 0.0000049 | 0.0387 |
| Taiwan | 2682 | 0.0331 | 0.0000012 | 0.0032 |

Next, we calculated the research performance parameters for individual countries (Table 1). The countries with the highest $e_p$ index values were Switzerland (0.15) and the USA (0.12), both of which were above the world average. All the other countries performed worse than the global average (< 0.10); the $e_p$ index varied from approximately 0.08 in the UK, Netherlands, Germany, and Sweden to less than 0.05 in South Korea, Japan, and Taiwan. The performance of the EU excluding the UK was similar to that of Canada and Australia. Among the four biggest continental EU countries (Germany, Spain, France, and Italy) Germany and France performed better than the EU excluding the UK. The probability of publishing a paper in the 0.01 percentile reflected the differences in the $e_p$ index, and the expected frequency of papers in the top 0.01% of the most cited publications, $P_{top\ 0.01\%}$, reflected both the differences in the $e_p$ index and in the number of publications. The $P_{top\ 0.01\%}$ indicator was 10-fold higher in the USA than in the EU excluding the UK and 5-fold higher than in the ERA.

To further investigate the dominant role of USA research, we calculated the performance parameters of collaborations between the USA and others countries (we actually studied co-authorship; Katz and Martin 1997). The results summarized in Table 2 reveal that any country collaborating with the USA substantially improved its $e_p$ index and paper probability of reaching the top 0.01 percentile. However, the increase in this probability was irregular; for example, it amounted 38-fold in Germany but only 2.7-fold in Switzerland. The increase of the probability was so high that although the number of collaborative papers was much lower than the number of domestic papers, the collaborative $P_{top\ 0.01\%}$ indicator was much higher than for domestic papers—with the exception of Switzerland. With reference to domestic papers, the collaboration between Switzerland and Germany increased 11-fold the probability for a paper from Germany to reach the 0.01 percentile but there was no increase for Switzerland's papers.

Table 2. Research performance parameters in hot medical topics, in seven cases of research collaborations between two countries. Publications in 2014.

| Countries | Number of publications | $e_p$ index | Paper probability for top 0.01% | $P_{top\ 0.01\%}$ |
|---|---|---|---|---|
| Germany and USA | 767 | 0.2001 | 0.001604 | 1.2303 |
| Canada and USA | 1059 | 0.1919 | 0.001356 | 1.4365 |
| Switzerland and USA | 206 | 0.1904 | 0.001315 | 0.2708 |
| Japan and USA | 696 | 0.1442 | 0.000433 | 0.3013 |
| China and USA | 3481 | 0.0990 | 0.000096 | 0.3343 |
| South Korea and USA | 774 | 0.0846 | 0.000051 | 0.0397 |
| Switzerland and Germany | 261 | 0.0812 | 0.000044 | 0.0114 |

Apparently, the USA also benefited from some of these collaborations, especially with Germany, Canada, and Switzerland, as they increased the domestic USA $e_p$ index; however, this is not the best explanation as we discuss below.

**Biochemistry and biotechnology**

In this case, the evolution of the number of publications over 30 years was substantially different from that observed in hot medical topics (compare Figures 2 and 3). The number of USA and ERA publications increased in the 1984–1994 period and remained constant or decreased after this time. Only Others retained a permanent growth throughout the 30-year period. It is worth noting that these searches were performed using research areas that were made up by a collection of journals. Therefore, the differential growth in the number of papers between areas is highly dependent on the inclusion of new journals in them and changes in topic preferences by researchers.

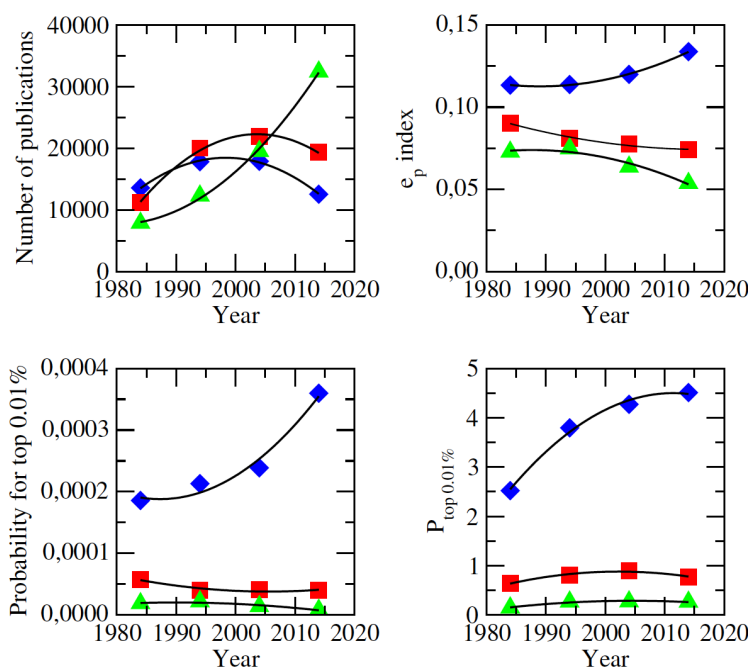

*Figure 3. Evolution of research performance in the USA, the European Research Area (ERA), and other countries in biochemistry and biotechnology over 30 years. Curves are drawn to guide the eye. Symbols: squares, the USA; diamonds, the ERA; triangles, other countries.*

Despite the differences in the evolution of the number of publications in the hot medical topics and biochemistry and biotechnology, the evolution of the $e_p$ index in these two research fields showed the similarity of small changes, positive evolution in the USA, and neutral or negative evolution in the other two geographical areas. The other size independent indicator, the probability that a paper reach the top 0.01 percentile, evolved as the $e_p$ index because it is mathematically linked to it. As observed in hot medical topics, also in biochemistry and biotechnology, the performance of the USA was already the best in 1984 and it increased over time.

Table 3. Research performance parameters in Biochemistry and Biotechnology research areas, in 15 selected cases. Publications in 2014; domestic counts.

| Country | Number of publications | $e_p$ index | Paper probability for top 0.01% | $P_{(top\ 0.01\%)}$ |
|---|---|---|---|---|
| USA | 12542 | 0.1323 | 0.0003068 | 3.847 |
| UK | 1795 | 0.1317 | 0.0003009 | 0.540 |
| Switzerland | 499 | 0.1219 | 0.0002210 | 0.110 |
| Netherlands | 652 | 0.1208 | 0.0002132 | 0.139 |
| Sweden | 470 | 0.1011 | 0.0001043 | 0.049 |
| Australia | 936 | 0.0938 | 0.0000775 | 0.073 |
| Germany | 2782 | 0.0827 | 0.0000468 | 0.130 |
| Canada | 1557 | 0.0771 | 0.0000354 | 0.055 |
| China | 10991 | 0.0740 | 0.0000301 | 0.330 |
| France | 1793 | 0.0677 | 0.0000211 | 0.038 |
| EU w/o UK | 15212 | 0.0662 | 0.0000192 | 0.292 |
| Italy | 1563 | 0.0534 | 0.0000081 | 0.013 |
| Spain | 1638 | 0.0505 | 0.0000065 | 0.011 |
| South Korea | 2379 | 0.0489 | 0.0000057 | 0.014 |
| Taiwan | 895 | 0.0418 | 0.0000031 | 0.003 |
| Japan | 3597 | 0.0375 | 0.0000020 | 0.007 |

The analysis of countries (Table 3) showed that the USA, the UK, Switzerland, and the Netherlands performed clearly better than the world average ($e_p$ index $> 0.1$). The performance by Sweden and Australia was approximately equal to the world average ($e_p$ index $= 0.1$), and all the other countries showed worse performances ($e_p$ index $< 0.1$). The EU excluding the UK showed rather a poor performance. As with hot medical topics, South Korea, Taiwan and Japan showed the lowest performances, which were considerably below the average global performance. The probability of publishing a paper in the 0.01 percentile reflected the differences in the $e_p$ index and the $P_{top\ 0.01\%}$ reflected both the differences in the $e_p$ index and in the number of publications. The $P_{top\ 0.01\%}$ indicator was 12-fold higher in the USA than in the EU excluding the UK and 4-fold higher than in the ERA.

Table 4. Research performance parameters in biochemistry and biotechnology research areas, in seven research collaborations between two countries. Publications in 2014

| Countries | Number of publications | $e_p$ index | Paper probability for top 0.01% | $P_{top\ 0.01\%}$ |
|---|---|---|---|---|
| Switzerland and USA | 112 | 0.2600 | 0.0045709 | 0.512 |
| Japan and USA | 285 | 0.1654 | 0.0007482 | 0.213 |
| Germany and USA | 409 | 0.1654 | 0.0007482 | 0.213 |
| Canada and USA | 403 | 0.1438 | 0.0004278 | 0.172 |
| Switzerland and Germany | 127 | 0.1303 | 0.0002887 | 0.037 |
| China and USA | 1403 | 0.1157 | 0.0001791 | 0.251 |
| South Korea and USA | 329 | 0.1062 | 0.0001273 | 0.042 |

Once again, collaboration with the USA substantially improved the research performance of all countries tested (Table 4); the increase in paper probability of reaching the top 0.01 percentile was very irregular. For example, it was more than 300-fold higher for Japan. For Switzerland and Germany the collaboration with the USA increased the top 0.01% probability by 21- and 16-fold, respectively. The collaboration between Switzerland-Germany increased 6-fold the top 0.01% probability of Germany but insignificantly the probability of Switzerland. Although the number of publications resulting from these collaborations was lower than the number of domestic papers, the cumulative frequency of publications in the top 0.01 percentile was higher in collaborations. Only in Germany did the improved research performance not compensate for the decreased number of publications.

Apparently, the USA also benefited from some of these collaborations, especially with Switzerland, as they increased the domestic USA $e_p$ index; however, as aforementioned, this is not the best explanation as we discuss below.

## Discussion

To quantify research performance in countries and institutions, we have used three parameters derived from citation distributions: the $e_p$ index, which reveals the *intrinsic efficiency* of the research system, the probability for a country publication to reach the global top 0.01 percentile, and the cumulative frequency of publications in this percentile. As explained previously, these parameters are mathematically calculated from the country's distribution of publications in the global publication percentiles based on citations. The $e_p$ index is percentile-independent while the other two parameters require a specific percentile to be selected (Rodríguez-Navarro and Brito 2018).

The rational behind the method of selecting a low percentile to estimate research performance is the assumption that the number of highly cited publications correlates with the number of discoveries or breakthroughs that a research system produces (Brito and Rodríguez-Navarro 2018 and references therein). This number of important discoveries or breakthroughs that boost science and breakthrough innovations is very low, as is the number of the highly cited papers that report these achievements. In consequence research performance has to be evaluated at a low percentile. The low number of breakthrough papers precludes counting these papers; however, the probability of its achievement can be calculated (Rodríguez-Navarro and Brito 2019). It is worth highlighting that the correlation between important achievements and the high number of citations does not imply that one of the parameters measures the other and that, therefore, it cannot be applied to low aggregation levels such as individual researchers or small groups (Allen et al. 2009; Ruiz-Castillo 2012; van Raan 2005). In conclusion, the probability that a paper published in a country reaches a low percentile equates to the country's probability of achieving important breakthroughs or discoveries of a similar frequency.

The selection of the top 0.01 percentile implies that in the topics and research areas studied, which produce approximately 150,000 and 80,000 annual publications, respectively, a total of 15 and 8 important discoveries or breakthroughs would be expected per year, which seems reasonable (Bornmann et al. 2018). Considering less important discoveries or breakthroughs a higher top percentile would be used, e.g., 0.1 or even 1.0. This less stringent percentile selection would reduce the country differences revealed by the probability and expected frequency at the top 0.01 percentile.

The hot medical topics we study here are quite different from the fast evolving technological topics we studied previously (Rodríguez-Navarro and Brito 2018), which belong to physical and chemical fields. However, in both cases, citation levels are similarly high, which reveals high interest of researchers for these research fields. This similarity may have been the reason why country research assessments in the two quite different scientific fields show many coincidences; however, for reasons we have not investigated they also show notable differences. The coincidence is the large advantage of the USA over the ERA, and the difference is that the Others are approaching to the USA in terms of the expected number of very highly cited publications in fast evolving technological but not in hot medical topics. Regarding the number of important discoveries and breakthroughs, our results suggest that the USA produces approximately 40% of them in fast evolving technological topics (Table 3 in Rodríguez-Navarro and Brito 2018) and approximately 80% in the hot medical topics (Figure 2). However, it is important to note that the USA remarkable high share of breakthroughs was obtained when publishing a quarter of the global number of papers.

The biochemical and biotechnological research areas studied here are not as highly cited as the hot medical topics; however, they are still highly cited, and the results are not largely different from those in hot medical topics (compare Figures 2 and 3).

Further studies are necessary to determine the relationship between R&D investments, the total number of publications, and the indicators used in this study. In any case, the large differences across countries in research performance attending to the probability of a paper reaching the top 0.01 percentile (Tables 1 and 3) strongly suggest that knowledge production depends more on research performance than on the total number of publications or R&D investments and capital. Therefore, the use of the last two parameters in econometric studies is necessarily misleading, because it is equivalent to considering a constant research performance across countries. We agree with the notion that "the benefits of scientific discovery have been heavy-tailed" (Press 2013, p. 822) but this and previous (Rodríguez-Navarro and Brito 2018) studies show that the frequency of the heavy-tailed discoveries vary enormously across countries. Therefore, it can be concluded that the benefits of research are likely to be highly variable across countries and that they can be high or low independently of R&D investments or capital, at least in economically advanced countries.

Here we have studied the research performance of the ERA, the EU, and other countries counting only domestic papers (i.e., excluding collaborations with external countries). This method does not measure the total scientific production of the research actors; however, in our opinion, it is the only method that can reveal the actual research performance level of a country or group of countries. Many arguments have been given in support of different counting methods for collaborative papers (Gauffriau 2017), but no method has been developed or can be developed to distribute unbalanced knowledge contributions among collaborating countries in unbalanced international collaborations (Zanotto et al. 2016). For example, the data in Tables 1–4 strongly suggest that considering collaborations between USA and China would mistakenly improve the research performance of Chinese researchers. It is worth noting that this apparent improvement of research performance that obtains a partner with a low research performance does not occur if the partner is a highly competitive country such as Singapore in fast evolving technological topics (Rodríguez-Navarro and Brito 2018).

Another issue is the apparent improvement in research performance that also occurs for the USA in collaborative papers; collaboration with Germany, Canada, or Switzerland increased the USA $e_\mathrm{p}$ index by 1.6-fold in hot medical topics (Tables 1 and 2) and 2.0-fold in the collaboration with Switzerland in biochemistry and biotechnology (Tables 3 and 4). However, this increase might not imply a real increase of the *intrinsic efficiency* of the USA in these collaborations. The most probable reason for this increase is that these international collaborations do not occur at random with all USA research institutions but occurs preferentially with top institutions in which the $e_\mathrm{p}$ index is much higher than the USA $e_\mathrm{p}$ index, which is a national average. For example, the $e_\mathrm{p}$ index of the Massachusetts Institute of Technology (MIT) or Harvard University in the two research fields studied here is almost 0.3 (unpublished results), which is similar to that measured in the aforementioned collaboration between USA and Switzerland. It is hard to believe that collaborations between MIT or Harvard University and Switzerland or Germany would improve the $e_\mathrm{p}$ index of MIT and Harvard University.

International collaborations (actually co-authorships; Katz and Martin 1997) are numerous (Leydesdorff et al. 2013; Wagner et al. 2015) and raise a complex problem in research evaluation and policy. This problem does not fall under the scope of this study and the conclusions above apply exclusively to the research topics and the research fields studied here.

On average, the probability that a paper reaches the top 0.01 percentile is much lower when the paper is published in EU continental countries than when it is published in the USA (approximately 10- and 16-fold in hot technological topics and biochemistry and biotechnology, respectively). Aside from the discussion of whether this countries' ratio accurately reflects the ratio for the achievement of discoveries and breakthroughs, the empirical fact cannot be denied and raises the question of why EU research is scarcely successful in publishing highly cited papers. The $e_p$ index in the most successful countries in continental Europe, Switzerland and the Netherlands, is similar to that of the USA; however, there are differences that raise some doubts about the comparison. Switzerland and the Netherlands are small countries with a low number of research universities that maintain a high level of research performance. In contrast, due to its size, the USA has many universities. In top USA universities the $e_p$ index is around 0.3 (unpublished results); however, at the end of the ranking the figures might be 100-fold lower.

From the point of view of the economy of a country, it seems logical that for a similar effect bigger countries will need higher numbers of scientific breakthroughs. To relate country size and research success the cumulative frequency for the top 0.01 percentile ($P_{top0.01\%}$) can be divided by the GDP or the number of inhabitants. Using the latter normalization in hot medical topics (Table 5), the first country is Switzerland, very prominent, and the second country is the USA; Netherlands and Sweden are the next two countries. In biochemistry and biotechnology, again the first two countries are Switzerland and the USA. The UK and Netherlands are the next two countries

Irrespective of the measurement method, it is evident that the four biggest continental EU countries keep a low research performance if the USA is taken as a reference, which raises a question about the causes. A key clue to answering this question might be that the greater differences occurs in fast evolving research topics (Bonaccorsi 2007; Rodriguez-Navarro and Narin 2018; Rodríguez-Navarro and Brito 2018; Sachwald 2015) that arouse greater interest in society and researchers. It is remarkable that in the WoS research areas "plant sciences" and "physiology" the USA and the EU are similarly successful. The $e_p$ index has not been calculated in these areas, but the similarity of the double rank plots (Figures 1 and 2 in Rodríguez-Navarro & Brito, 2018a) allows the prediction of very similar $e_p$ index values. This situation is puzzling, suggesting differences in researchers' motivations.

Table 5. $P_{top\ 0.01\%}$ per million inhabitants across countries

| Country | Hot medical topics | Biochem & Biotechnol |
|---|---|---|
| Switzerland | 0.04387 | 0.01379 |
| USA | 0.01743 | 0.01202 |
| Netherlands | 0.00444 | 0.00818 |
| Sweden | 0.00429 | 0.00490 |
| Germany | 0.00249 | 0.00159 |
| UK | 0.00235 | 0.00818 |
| Australia | 0.00154 | 0.00302 |
| Canada | 0.00148 | 0.00149 |
| France | 0.00132 | 0.00058 |
| Spain | 0.00065 | 0.00025 |
| Italy | 0.00062 | 0.00022 |
| South Korea | 0.00062 | 0.00027 |
| Japan | 0.00030 | 0.00006 |
| China | 0.00017 | 0.00024 |
| Taiwan | 0.00013 | 0.00011 |

# Conclusions

Citation distributions can be used to reveal the scientific competence and efficiency of countries. The $e_p$ index reveals the efficiency of the research system in scaling up from less cited papers to highly cited papers with reference to global papers. This efficiency can also be quantified through the probability of publishing a paper surpassing a certain high level of citations, which is a rare event like breakthroughs and discoveries. The high variability of this probability across economically advanced countries indicates that successful research depends on intrinsic characteristics of research systems that are currently unknown. A gaming simile would be that, for an unknown reason, throwing a seven using two dice had a probability of 0.17 in the USA and 10-fold less in the EU.

If researcher curiosity is the driving force behind discoveries, it seems that researchers in different countries have curiosities of different scientific and societal relevance. Furthermore, our 30-year study of the evolution of the scientific performance of the USA and the EU suggests that scientific competence and efficiency is an intrinsic characteristic of old research systems that does not change easily. In any case, a Mathew effect seems to exist, which makes the strong USA research to evolve stronger while the weak EU research undergoes minimal positive or negative fluctuations.

**Acknowledgements**

This work was supported by the Spanish Ministerio de Economía y Competitividad grant number FIS2017-83709-R

Brito, R and Rodríguez-Navarro, A (2018), 'Research assessment by percentile-based double rank analysis', *Journal of Informetrics,* 12, 315-29.

Dewar, R D and Dutton, J E (1986), 'The addoption of radical and incremental innovations: An empirical analysis', *Management Science*, 32, 1422-33.

Gauffriau, M (2017), 'A categorization of arguments for counting methods for publication and citation indicators', *Journal of Informetrics,* 11, 672-84.

Katz, J S and Martin, B R (1997), 'What is reserach collaboration?', *Research Policy,* 26, 1-18.

Kuhn, T (1970), *The structure of scientific revolutions* (Chicago: University of Chicago Press).

Leydesdorff, L, et al. (2013), 'International collaboration in science: The global map and the network', *El Profesional de la Información*, 22, 87-94.

Martin, B R and Irvine, J (1983), 'Assessing basic research. Some partial indicators of scientific progress in ratio astronomy', *Research Policy,* 12, 61-90.

Press, W H (2013), 'What's so special about science (and how much should we spend on it?)', *Science,* 342, 817-22.

Rodriguez-Navarro, A and Narin, F (2018), 'European paradox or delusion-Are European science and economy outdated?', *Science and Public Policy,* 45, 14-23.

Rodríguez-Navarro, A (2012), 'Counting highly cited papers for university research assessment: conceptual and technical issues', *PLoS One,* 7(10), e47210.

Rodríguez-Navarro, A and Brito, R (2018), 'Technological research in the EU is less efficient than the world average. EU research policy risks Europeans' future', *Journal of Informetrics,* 12, 718-31.

--- (2019), 'Probability and expected frequency of breakthroughs – basis and use of a robust method of research assessment', *Scientometrics,* 119, 213-235.

Röhrig, B, et al. (2009), 'Types of study in medical research. Part 3 of a series on evaluation of scientific publications', *Deutsches Ärzteblatt International* 106, 262-68.

Ruiz-Castillo, J (2012), 'The evaluation of citation distribution', *SERIEs,* 3, 291-310.

Sachwald, F (2015), *Europe's twin deficits: Excellence and innovation in new sectors* (Luxembourg: Publications Office of the European Union).

Salter, A J and Martin, B R (2001), 'Economic benefits of public funded basic research: a critical review', *Research Policy,* 30, 509-32.21